\documentclass[epj]{svjour}
\usepackage{graphics}

\newcommand{\pbar}   {\mbox{$\bar{\mathrm p}$}}

\newcommand{\jpsi}   {\mbox{$\mathrm{J/\psi}$}}

\begin{document}
\title{W Mass and Properties}
\author{Mark Lancaster (on behalf of the CDF and {D\O} collaborations)}
\institute{UCL, Department of Physics and Astronomy, 
Gower Street, London, UK, WC1E 6BT.}
\date{Received: date / Revised version: date}
%
\abstract{
Precise measurements of the mass and width of the W boson are
sensitive to radiative corrections and can be used to place limits on
new physics beyond the Standard Model and validate the consistency of
the model.  In particular, the W boson mass constrains the mass of
the, as yet unobserved, Higgs boson and the width can be used to place
limits on the existence of new particles that couple to the W. Results
are presented from p{\pbar} collisions recorded by the CDF and {D\O}
experiments at the Fermilab Tevatron collider, operating at a centre
of mass energy of 1.96 TeV. The uncertainty on the W mass is
determined to be 76 MeV by CDF and the width, by {D\O}, to be 2011 $\pm$
90 (stat.) $\pm$ 107 (syst.) MeV.
} 
\maketitle
\section{Introduction}
\label{intro}
The world's largest sample of W bosons is presently being analysed by
the CDF and {D\O} collaborations.
The results presented here are based on an integrated luminosity of
$\sim 200{\rm pb}^{-1}$, accumulated in 2002-2003; which is a factor
of two larger than used in the previously published
results~\cite{RUN1_RESULTS}. Results on the W production cross
section, angular distribution and couplings to other gauge bosons have
been presented at this conference~\cite{WPROPS_HCP2005}. In this talk
results on the W boson mass and width will be presented. The
results are important in verifying the consistency of the Standard
Model, placing limits on new physics, and in determining the mass of
the Higgs boson.

\section{CDF W Mass Measurement}
\label{sec:CDF_WMASS}
\begin{figure}
\resizebox{0.5\textwidth}{!}{
\includegraphics{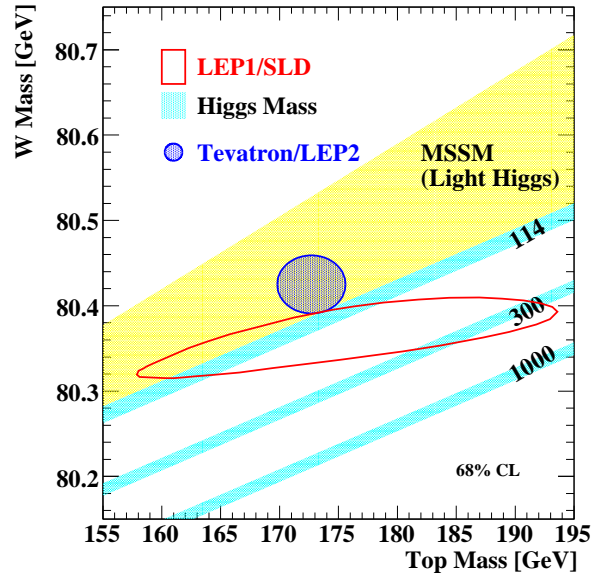}}
\caption{The predicted W boson and top quark mass in the Standard
Model for three Higgs masses (114 - the lower limit from LEP direct
searches, 300 and 1000 GeV) and in the MSSM extension to the Standard
Model. The present constraint from the Tevatron top and W mass and
LEP2 W mass measurements are shown. The indirect constraint from
precision electroweak measurements at LEP1 and SLD is also shown.}
\label{fig:1}      
\end{figure}

At tree level, the mass of the W boson is determined by the mass of
the Z boson (which has been very precisely measured at
LEP~\cite{LEP_ZMASS}) and the electromagnetic and weak coupling
constants. Beyond tree level, it is subject to radiative corrections
which depend on the masses of all the particles the W can couple to.
The largest contribution comes from the top quark and there is a weak
dependence on the mass of the Higgs boson. Precision
measurements of the W boson mass, in conjunction with a top quark mass
measurement~\cite{TOPMASS_HCP2005}, can therefore be used to constrain the mass
of the Higgs boson and other more exotic particles e.g. those
predicted by super-symmetric (SUSY) models.  This is shown in
figure~\ref{fig:1}, which shows the predicted variation of the W and
top masses for three choices of the Higgs mass and the region favoured
by the minimal SUSY extension to the Standard Model (MSSM) with a
light Higgs boson. In general scenarios with a light Higgs and SUSY particles
tend to raise the mass of the W boson.

At hadron colliders the W mass is measured in the electron and muon
decay channels since these channels can be identified with high
efficiency and with little background contamination. However, with
these decay modes there is an accompanying neutrino whose momentum can
only be inferred through momentum conservation in the transverse
plane. As such the mass of the W boson has to be determined from a
measurement of the mass using transverse momentum components only.  It
is not possible to have a simple functional form, in terms of the true
W mass, for this transverse mass owing to the effects of the varying
parton-parton centre of mass energy, and the detector acceptance and
resolution. Templates of the transverse mass distribution after a full
simulation of the physics and the detector are therefore generated at
various W mass values and the W mass is ultimately obtained from a
likelihood comparison of the data with these templates. Events are
generated using the NLO QCD generator RESBOS~\cite{RESBOS} and the
effect of photon radiation from the decay charged leptons is taken
from the WGRAD~\cite{WGRAD} calculation. This calculation only
simulates the emission of a single photon and the uncertainty in the W
mass arising from not including further emissions has been estimated
to be 15 (20) MeV in the electron (muon) channel respectively. 
Owing to the similarity in the production mechanism between W and Z
bosons, it is possible to predict the W transverse momentum
distribution from a measurement of the Z transverse momentum
distribution using the decay leptons. The uncertainty in the W
transverse momentum, due
to the finite statistics of the calibrating Z sample, results in a 15
MeV uncertainty in the W mass. The uncertainty in the angular
distribution of the W bosons, arising from uncertainties in the parton
distribution functions (PDFs) is determined using the CTEQ6~\cite{CTEQ6} 
and MRST~\cite{MRST} PDFs and is determined to be 15 MeV.

A key aspect of the measurement of the W mass is the determination of the
momentum and energy scale of the charged leptons from the tracking
detectors and the calorimeter. For the muons, the momentum scale is
set using measurements of the \jpsi\ and Upsilon masses. For the
electrons, the energy scale is set by requiring the energy scale to
match the momentum scale (already set from the {\jpsi}).  Both these
determinations require a very detailed simulation of the photon
radiation in the passive material, both in terms of simulating all
possible physics processes but also in the composition and location of
the material. The scale uncertainties are determined to be 70 and 25
MeV for the electron and muon channels respectively. The resolution of
the energy and momentum measurements are taken from a fit to the width
of the Z invariant mass distributions and the finite Z statistics result in a 15
MeV W mass uncertainty from this source for both channels.

\begin{figure}
\resizebox{0.5\textwidth}{!}{
\includegraphics{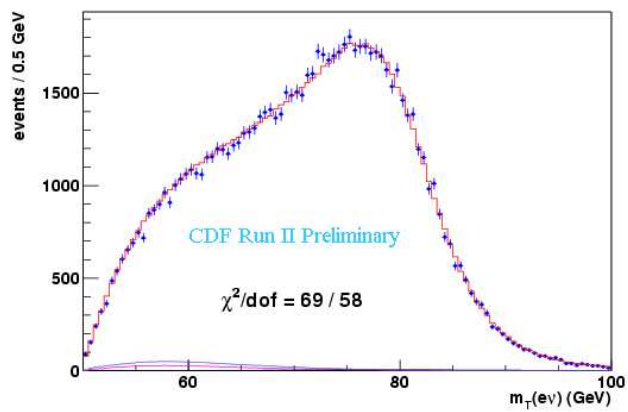}}
\caption{The transverse mass distributions of the W $\rightarrow$
 e$\nu$ sample used to extract the W mass.}
\label{fig:2}      
\end{figure}

In order to determine the neutrino momentum, through momentum
conservation in the transverse plane, it is necessary to have a
simulation of the underlying event, concurrent minimum bias event
and the initial state QCD radiation. These components cannot be
accurately modelled using a standard Monte Carlo event generator and
are instead parameterised by fitting a model to real minimum bias and
Z events; whose characteristics with regard the underlying event and
QCD radiation are expected to be very similar to W
events. Uncertainties in this model arise from the finite statistics
of the Z sample and from biases induced by the differing selection
criteria and acceptance of the Z and W events e.g. Z events are
selected with both leptons in the central detector region, whereas in
W events there can be no such constraint on the direction of the
neutrino. These uncertainties contribute a 50 MeV uncertainty in the W
mass in both channels.  The two largest sources of background : W to
$\tau$ decays with subsequent $\tau$ decay to e$\nu\nu$ or $\mu\nu\nu$ and
Z events where the second charged lepton escapes detection can be
accurately simulated and the level of background (typically $\sim$
5\%) can be reliably estimated from the simulation. Backgrounds from
QCD processes, cosmic rays and decay in flight Kaons cannot be accurately
simulated and estimates of the transverse mass distributions from
these sources are taken from the data by relaxing the selection cuts
to provide background rich samples. Uncertainties in the level and
shape of the background distributions contribute $\sim$ 20 MeV to the
W mass uncertainty. The complete list of systematic uncertainties for
the CDF W mass analysis are shown in table~\ref{table:syst}. The total
combined error, after taking into account correlations between the two
channels, is 76 MeV. This is better than the previously published CDF
W mass which had an uncertainty of 79 MeV. This systematic error
analysis is a preliminary one and it expected to be reduced before
publication. The transverse mass distributions of the electron sample
used to determine the W mass is shown in figure~\ref{fig:2}.

\begin{table}
\caption{Systematic and statistical uncertainties (in MeV) for the CDF W mass analysis}
\label{table:syst}       
\begin{tabular}{lll}
\hline\noalign{\smallskip}
Error Source & W $\rightarrow$ e$\nu$ & W $\rightarrow$ $\mu\nu$   \\
\noalign{\smallskip}\hline\noalign{\smallskip}
Statistics                            & 45 & 50  \\
Production model \& decay             & 30 & 30  \\
Charged lepton scale \& resolution    & 70 & 30  \\
Backgrounds                           & 20 & 20  \\
Recoil scale \& resolution            & 50 & 50  \\ 
Total                                 & 105 & 95 \\
\noalign{\smallskip}\hline
\end{tabular}
\end{table}

\section{W Width measurement}
\label{sec:2}
\begin{figure}
\resizebox{0.5\textwidth}{!}{
\includegraphics{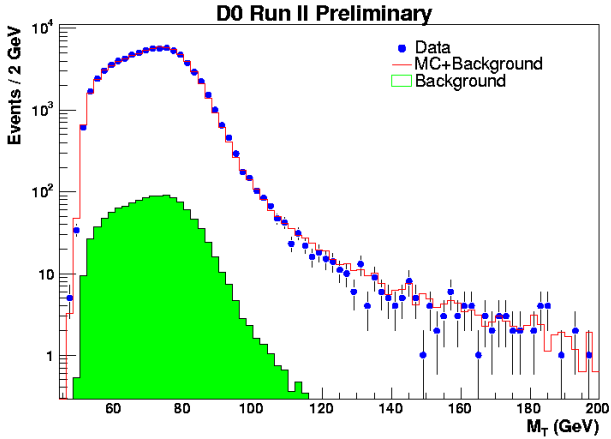}}
\caption{The transverse mass distributions of the W $\rightarrow$
 e$\nu$ sample used to extract the W width by {D\O}.}
\label{fig:3}      
\end{figure}

As seen in figure~\ref{fig:2} the W transverse mass distribution has a sharp
edge close to the value of the W mass. However owing to the finite
width of the W boson, it is also possible for events to be measured
with transverse mass values higher than the mass of the W boson. From
a likelihood fit to the transverse mass distribution in the $100 < m_T
< 200$ GeV, it is therefore possible to determine the W width. However
events in the high transverse mass region can also arise due to the
finite resolution of the detector and so a detailed understanding and
modelling of resolution effects is a vital component of this analysis
and indeed dominates the systematic uncertainty for the measurement. 
Using 177 pb$^{-1}$ of data, and 625 W $\rightarrow$ e$\nu$ events in
the high transverse mass region, {D\O} have determined the W width to
be 2011 $\pm$ 93 (stat.) $\pm$ 107 (syst.) MeV; which
agrees well with the Standard Model prediction of 2099 $\pm$ 3
MeV~\cite{SM_WIDTH}. The transverse mass distribution of the
W $\rightarrow$ e$\nu$ events used by {D\O} to determine the W width 
are shown in figure~\ref{fig:3}. 

\section{Future measurements}
\label{sec:3}
The analyses presented here have been based on an integrated
luminosity of $\sim$ 200~pb$^{-1}$. At the time of this conference the
Tevatron had passed the 1~fb$^{-1}$ milestone and the next set of W
width and mass measurements are expected to be based on datasets of
1-2~fb$^{-1}$. In these analyses the limiting factor in precision will
be systematic and not statistical. The systematic
uncertainties arising from PDFs and QED radiative corrections are
likely to the limiting source of error in these analyses. At present
these two sources contribute $\sim$ 25 MeV to the W mass uncertainty
and this is common to the two experiments. Further developments in
parton fitting (additional d/u data from HERA and a more sophisticated
error analysis) and the provision of a fast generator that incorporates both 
NLO QED (i.e $\cal{O}$($\alpha^2$)) {\it and} NLO QCD are likely to
be needed if this 25 MeV uncertainty is to be reduced. The
expectations are that with a 2~fb$^{-1}$ dataset the Tevatron
experiments will produce a W mass with a combined uncertainty of 20-30
MeV and a width uncertainty of 35~MeV. These uncertainties will
surpass those from LEP2; furthermore each experiment will have more precise
measurements than any single LEP experiment.

\section{Acknowledgements}
\label{sec:4}
I'd like to thank the organisers of the Hadron Collider symposium for
a supremely well organised, stimulating and enjoyable conference and the Royal Society for providing funding that enabled me to present these results.

\end{document}